\documentclass[twocolumn]{aastex62} 

\usepackage{xspace}
\usepackage{color}
\usepackage{rotating}
\usepackage{comment}
\usepackage{amsmath}
\usepackage{subfigure}
\usepackage[ampersand]{easylist}
\usepackage[version=4]{mhchem} 
\usepackage[normalem]{ulem}    
\usepackage[nameinlink]{cleveref}

\usepackage{lineno}

\PassOptionsToPackage{hyphens}{url}\usepackage{hyperref}

\newcommand{\chimera}{\texttt{CHIMERA}\xspace}

\newcommand{\pandexo}{\texttt{PandExo}\xspace}
\newcommand{\pandeia}{\texttt{Pandeia}\xspace}

\def\gsim{~\rlap{$>$}{\lower 1.0ex\hbox{$\sim$}}}
\def\lsim{~\rlap{$<$}{\lower 1.0ex\hbox{$\sim$}}}
\newcommand{\emcee}[0]{\texttt{emcee}\xspace}

\newcommand{\jwst}[0]{JWST\xspace}

\defcitealias{Stevenson2020}{S20}

\begin{document}

\bibliographystyle{apj}

\title{Retrieving Exoplanet Atmospheres using Planetary Infrared Excess: \\Prospects for the Nightside of WASP-43 b and other Hot Jupiters}

\shorttitle{ExoPIE}
\shortauthors{Lustig-Yaeger et al.}

\correspondingauthor{Jacob Lustig-Yaeger}
\email{Jacob.Lustig-Yaeger@jhuapl.edu}

\author[0000-0002-0746-1980]{Jacob Lustig-Yaeger}
\affiliation{Johns Hopkins University Applied Physics Laboratory, Laurel, MD 20723, USA}
\affiliation{NASA NExSS Virtual Planetary Laboratory, Box 351580, University of Washington, Seattle, Washington 98195, USA}

\author[0000-0002-7352-7941]{Kevin B. Stevenson}
\affiliation{Johns Hopkins University Applied Physics Laboratory, Laurel, MD 20723, USA}

\author[0000-0002-4321-4581]{L. C. Mayorga}
\affiliation{Johns Hopkins University Applied Physics Laboratory, Laurel, MD 20723, USA}

\author[0000-0001-7393-2368]{Kristin Showalter Sotzen}
\affiliation{Johns Hopkins University Applied Physics Laboratory, Laurel, MD 20723, USA}
\affiliation{Johns Hopkins University, 3400 N. Charles Street, Baltimore, MD 21218, USA}

\author[0000-0002-2739-1465]{Erin M. May}
\affiliation{Johns Hopkins University Applied Physics Laboratory, Laurel, MD 20723, USA}

\author[0000-0003-1629-6478]{Noam R. Izenberg} 
\affiliation{Johns Hopkins University Applied Physics Laboratory, Laurel, MD 20723, USA}

\author[0000-0001-8397-3315]{Kathleen Mandt} 
\affiliation{Johns Hopkins University Applied Physics Laboratory, Laurel, MD 20723, USA}

\begin{abstract}
To increase the sample size of future atmospheric characterization efforts, we build on the planetary infrared excess (PIE) technique that has been proposed as a means to detect and characterize the thermal spectra of transiting and non-transiting exoplanets using sufficiently broad wavelength coverage to uniquely constrain the stellar and planetary spectral components from spatially unresolved observations. 
We performed simultaneous retrievals of stellar and planetary spectra for the archetypal planet WASP-43b in its original configuration and a non-transiting configuration to determine the efficacy of the PIE technique for characterizing the planet's nightside atmospheric thermal structure and composition using typical out-of-transit JWST observations.  
We found that using PIE with JWST should enable the stellar and planetary spectra to be disentangled with no degeneracies seen between the two flux sources, thus allowing robust constraints on the planet's nightside thermal structure and water abundance to be retrieved. The broad wavelength coverage achieved by combining spectra from NIRISS, NIRSpec, and MIRI enables PIE retrievals that are within 10\% of the precision attained using traditional secondary eclipse measurements, although mid-IR observations with MIRI alone may face up to $3.5\times$ lower precision on the planet's irradiation temperature.  
For non-transiting planets with unconstrained radius priors, we were able to identify and break the degeneracy between planet radius and irradiation temperature using data that resolved the peak of both the stellar and planetary spectra, thus potentially increasing the number of planets amenable to atmospheric characterization with JWST and other future mission concepts. 
\end{abstract}


\section{Introduction} 
\label{sec:intro}

While transiting planets offer a powerful means to characterize their atmospheres using transit and eclipse spectroscopy, most exoplanets do not transit their host stars as seen from Earth.  Similarly, the study of directly-imaged exoplanets is currently limited to the youngest, hottest planets at wide separations.  As a result, new observational, technological, or theoretical advances that can separate planets from stars are required to enable the atmospheric study of the large population of non-transiting exoplanets orbiting main sequence stars.

Transiting exoplanets have a reference point \textit{in time} (secondary eclipse) during which the planetary flux is blocked, allowing the stellar and planetary flux to be separated using \textit{temporally resolved} measurements (i.e., time-series observations). Directly-imaged exoplanets can be observed with a coronagraph or starshade so that the planet's flux can be \textit{spatially resolved} from that of the host star. Recently, \citet{Stevenson2020} (henceforth \citetalias{Stevenson2020}) proposed that the sufficiently broad wavelength coverage provided by the James Webb Space Telescope \citep[JWST;][]{Gardner2006, Beichman2014, Kalirai2018} will allow the planet and star to be \textit{spectrally resolved}. By leveraging the fact that planets and stars of significantly different temperature have separable spectral energy distributions (SEDs), it is possible to fit the two flux sources simultaneously and thereby recover the planetary infrared excess (PIE). 

\citetalias{Stevenson2020} demonstrated that the PIE technique could be used to simultaneously infer the radius and brightness temperature of non-transiting exoplanets by acquiring simultaneous, broad-wavelength spectra, and discuss how additional planet atmospheric characteristics (such as composition and thermal structure) could be retrieved. \citetalias{Stevenson2020} also presented several potentially-viable applications of the PIE technique, including efficiently studying atmospheric dynamics by obtaining sparsely-sampled phase curve observations, measuring nightside temperatures with transit observations, constraining the radius of directly-imaged planets, and searching for biosignatures in planets orbiting the nearest M-dwarf stars.  Of these applications, measuring the planet's nightside temperature will be the most straightforward to validate with JWST.

\citet{Arcangeli2021} recently applied a similar approach to the PIE technique using sparsely-sampled phase curve observations of WASP-12b from the Hubble Space Telescope's WFC3 instrument.  Using the measured stellar spectrum in eclipse, they placed constraints on the planet's relative spectrum (up to an additive constant) at quadrature phase and, by fitting the slope of the emission spectrum, were able to estimate WASP-12b's brightness temperature.  Attempts to repeat this experiment for the planet's nightside using out-of-transit data were unsuccessful due to a significant change in the position of the spectrum on the detector that lead to relatively large systematics.  Due to JWST's stable location near L2, its pointing repeatability is expected to be significantly better than that of HST \citep[which orbits the Earth,][]{Stevenson2019a}.

Based on the novel results from \citet{Arcangeli2021}, it is clear that the PIE technique offers an invaluable opportunity to study both transiting and non-transiting thermally bright exoplanets.  The transit probability for a hot Jupiter orbiting a Sun-like star is $\sim10\%$, thus for every HD~209458b or HD~189733b-like transiting planet, there are ten times as many non-transiting exoplanets orbiting similarly bright stars.  However, before pointing JWST at the nearest known non-transiting hot Jupiter, numerous potential obstacles must be considered in order to demonstrate that the method is not simply {\em PIE in the sky}. 
In this paper, we confront the following questions regarding the efficacy of the PIE technique for studying the nightside of the hot Jupiter WASP-43b with JWST:  
\begin{enumerate}
    \item Can a planet's atmospheric thermal structure and composition be retrieved using the PIE technique? 
    \item How degenerate are the stellar and planetary parameters in simultaneous fits to an unresolved spectrum? 
    \item What are the optimal JWST instruments and exposure times for PIE observations? 
    \item What is the prospect for PIE on non-transiting exoplanets?  
    \item What impact does exozodiacal dust have on the retrieved parameters?
\end{enumerate}

In Section \ref{sec:methods}, we detail novel modifications to a standard exoplanet atmospheric retrieval framework to enable PIE retrievals. In Section \ref{sec:results}, we present retrieval results using the PIE technique for WASP-43b with JWST, touching on the sensitivity to wavelength range and exposure time for transiting and non-transiting exoplanets. In Section \ref{sec:discussion}, we discuss our findings in the broader context of the field and conclude with a roadmap for the continued maturity of the PIE technique for exoplanet characterization. 

\section{Methods} 
\label{sec:methods}

\subsection{Retrieval Model} 
\label{sec:methods:model} 

We developed a modified version of the CaltecH Inverse ModEling and Retrieval Algorithms (\chimera) code \citep{Line2013a, Line2014} to operate using the PIE technique. 
\chimera is a flexible, open-source, and well-established retrieval framework used to infer the atmospheric composition of exoplanets from emission and transmission spectroscopy measurements in the optical to mid-infrared, for a wide variety of planet types (terrestrial to ultra-hot-Jupiter) given arbitrary molecular abundances (constant-with-altitude or variable with altitude), thermal structures, and cloud/aerosol properties (droplet sizes, single scatter albedo/asymmetry parameter, vertical distribution, \citep{Mai2019}). The core radiative transfer scheme accounts for full multiple scattering of outgoing thermal radiation utilizing the methods outlined by \citep{Toon1989}. 
We incorporate correlated-K \citep[``resort-rebin'', ][]{Amundsen2017} opacities by generating pre-computed, high-resolution cross-section grids sourced from a variety of databases including ExoMol \citep{Tennyson2016, Tennyson2018}, utilizing the EXOCROSS routine \citep{Yurchenko2018}, and HITRAN/HITEMP \citep{Rothman2010}, via the HAPI routine \citep{Kochanov2016HAPI}. 
The transmission and emission routines have been validated on numerous occasions against other codes \citep{Line2013c, Line2013a, Morley2015}, and used to analyze brown dwarf observations \citep{Line2014b, Line2015, Line2017}. 

The most significant change required to implement PIE into the standard \chimera emission forward model is to use the absolute flux from the system, $F_{sys}$. This differs from the standard emission spectroscopy approach, which models the wavelength-dependent secondary eclipse depth as the planet-to-star flux ratio, $\Delta F = F_p/F_s$. In this case, we define the system flux as the sum of all major flux sources in the exoplanetary system: 
\begin{equation}
\label{eqn:system_flux}
    F_{sys} = F_{s} + F_{p} + F_{\epsilon}, 
\end{equation}
where $F_s$ is the stellar flux, $F_p$ is the planet flux, and $F_{\epsilon}$ is a catch-all term to account for any residual astrophysical flux source in the field (such as an additional, cooler planet or exozodiacal dust). For simplicity we set $F_{\epsilon} = 0$ and focus on joint modeling of the stellar and planetary flux. We explore the ramifications of $F_{\epsilon} = F_{ez}$ due to exozodiacal light in Section \ref{sec:results:zodi}.  

\subsubsection{Planetary Flux Model} 

We use the standard \chimera emission spectrum forward model presented by \citet{Line2013a, Line2014} and \citet{Batalha2018} with no modifications to the physics and chemistry. We employ the free chemistry (henceforth ``free'') 
forward model in our exploration of the PIE technique, which allows for the abundances of the gases in the state vector to take on any (physical, even if chemically implausible) values and assumes evenly mixed volume mixing ratio vertical profiles. 
As such, we directly fit for the (log) gas abundances. 
Although we performed tests using the ``chemically consistent'' \chimera forward model \citep[e.g.,][]{Kreidberg2015}, which assumes thermochemical equilibrium gas vertical profiles, our results were generally consistent with those acquired with the ``free'' model and have been omitted for clarity and brevity.  

For the temperature-pressure (TP) profile, we use the three-parameter, analytic radiative equilibrium model from \citet{Guillot2010} \citep[see also, ][]{Parmentier2014, Line2013a, Mai2019}. The irradiation temperature $T_{irr}$, the infrared opacity $\log(\kappa_{IR})$, and the single channel visible to IR opacity ratio $\log(g_1)$ are treated as free parameters, and the internal temperature $T_{int}$ is fixed at 200 K.  

Bringing together all of the physical, atmospheric, and thermal parameters that define our planet model, the planet flux is given as, 
\begin{align}
    F_p =& F_p(\theta_p) \\ 
    \theta_{p} = \{  & R_p, T_{irr}, \log(\kappa_{IR}), \log(g_1), \log(\ce{H2O}) \nonumber \\
                   & \log(\ce{CH4}), \log(\ce{CO}), \log(\ce{CO2}), \log(\ce{NH3}) \} . 
\end{align}

\subsubsection{Stellar Flux Model}

We use Phoenix stellar models as the basis for our stellar flux model \citep{allard2003model, allard2007k, Allard2012}. We access the stellar grids using the \texttt{pysynphot} code \citep{STScI2013} and linearly interpolate in effective temperature $T_{eff}$, surface gravity $\log(g)$, and metallicity $\log([M/H])$ between computed stellar models. We scale the stellar flux using the stellar distance, $d$, and stellar radius, $R_s$, so as to reproduce stellar flux observations at Earth. This results in a five parameter stellar flux model that is parameterized via the following state vector, 
\begin{align}
    F_{s} &= F_{s}(\theta_s) \\
    \theta_s &= \{ T_{eff}, \log(g), \log([M/H]), d, R_s \}.  
\end{align}

\subsubsection{Inverse Model} 

We use the \emcee \citep{Foreman-Mackey2013} implementation of the affine-invariant Markov Chain Monte Carlo (MCMC) ensemble sampler from \citet{Goodman2010}. For each simulation we set the number of walkers equal to ten times the dimensionality of the model and randomly draw their initial positions from an estimate of the posterior variance obtained through Levenberg-Marquardt least squares minimization \citep{More1978levenberg}. We iteratively draw $N$ samples until $N \gsim 50 \tau$, where $\tau$ is the integrated autocorrelaion time. We performed a variety of tests using the \texttt{dynesty} Nested Sampling algorithm \citep{Skilling2004, Speagle2020}, but found that it converged inefficiently due to the vast dynamic range in contribution to the likelihood between the stellar and planetary parameters. Therefore, our reported results were obtained exclusively using MCMC. 

\subsection{Synthetic JWST Observations} 
\label{sec:methods:obs}

We use \pandexo to simulate observations of WASP-43b with \jwst \citep{Batalha2017b, Pandexo2018}. \pandexo uses the official \jwst exposure time calculator (ETC), \pandeia \citep{Pontoppidan2016}, to perform noise modeling for exoplanet observations. In this work, we focus exclusively on observations in the 0.6--12 {\micron} wavelength range using NIRISS SOSS, NIRSpec G395, and MIRI LRS because these instruments provide complete wavelength coverage over the stated range without saturating. We bin our simulated \jwst data to a constant resolving power of $\mathcal{R}=100$ for our retrieval analyses. 

To study the efficacy of the PIE technique, we simulated high-resolution model spectra using the forward model described in Section \ref{sec:methods:model} before implementing noise from \pandexo. Furthermore, we do not add random Gaussian jitter to the data.  Although these assumptions yield idealized observations that, by design, will be capable of being well-fit by our nominal model, this exercise is common practice in the exoplanet retrieval literature \citep[see][for more details]{Feng2018} and allows us to characterize bias and degeneracies in the retrieval procedure, which are of keen interest in this study. Although we calculate the precision of our synthetic spectra using only \pandexo noise calculations, we considered the effect of absolute flux calibration accuracy in \autoref{sec:appendix:calibration} as it may complicate the study of broad wavelength spectra constructed using multiple instruments with offsets in absolute flux between them. 

The first two columns of Table \ref{tab:results_free} provide the default choice of model parameters that were used to simulate the JWST observations. 
We used Gaia DR2 data for the stellar parameters \citep{Gaia2018, Bailer-Jones2018}. For the planetary parameters, we used gas abundances and a TP profile that are consistent with the current estimates from retrievals of WASP-43b's transmission spectrum \citep{Kreidberg2014}. However, one notable departure from this convention is that we set the irradiation temperature ($T_{irr}$) to 1000 K to be representative of the nightside of this hot Jupiter \citep{Beatty2019}.  The third column of Table \ref{tab:results_free} shows the Bayesian prior probability distributions used throughout this investigation, unless otherwise stated. 

\section{Results} 
\label{sec:results}

\begin{figure*}[!t]
\centering
\includegraphics[width=0.95\textwidth]{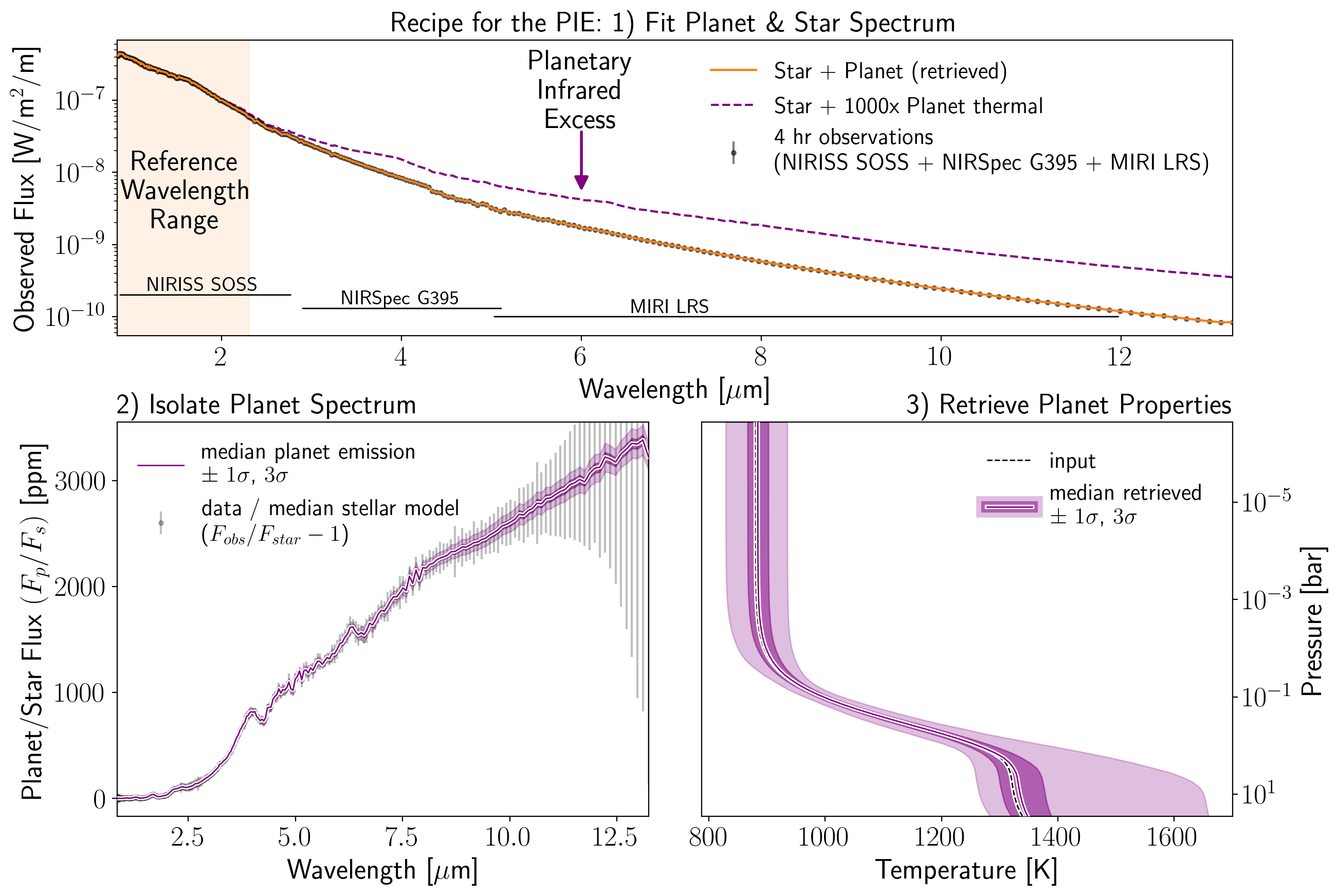}
\caption{\textit{Top:} Figure depicting the concept of PIE for the WASP-43 planetary system. The observed stellar plus nightside planetary flux ({$\sim$}1000\,K) is shown simulated across the JWST wavelength range, enabling a retrieval of WASP-43b's nightside atmospheric properties. The planet thermal emission is shown enhanced by a factor of 1000\,K for visual clarity, although it is \textbf{not} enhanced in the lower panels or for the actual analysis. Using data from the reference wavelength range ($<2.3$~{\micron}) to calibrate the stellar model, we can isolate and infer the planet's infrared excess by simultaneously fitting the stellar and planet spectra. 
\textit{Lower Left:} The planet's thermal emission is shown relative to the median retrieved stellar model with 1$\sigma$ and 3$\sigma$ credible regions. 
\textit{Lower Right:} Retrieved constraints on WASP-43b's nightside temperature-pressure profile.  There is no retrieved bias relative to the input TP profile.} 
\label{fig:pie_spectrum}
\end{figure*}

\Cref{fig:pie_spectrum} shows a conceptual demonstration of the PIE technique. The top panel of \Cref{fig:pie_spectrum} depicts a model spectrum of the WASP-43 system, where the observed flux is a combination of the stellar ({$\sim$}4400\,K) and planetary nightside (${\sim}$1000\,K) spectra. We include simulated 1$\sigma$ uncertainties on the observed flux, corresponding to a 4-hour exposure with each JWST instrument mode considered, but the error bars are much smaller than the point size. As in \citet{Stevenson2020}, we show the PIE signal enhanced by a factor of $1000 \times$ to aid in visualizing the contribution to the total system flux from the planet as a function of wavelength. However, we emphasize that this enhancement is not used in any part of the analysis. Crucially, the planet provides negligible flux compared to the star at short wavelengths near the peak of the stellar spectral energy distribution (SED), which yields a ``reference wavelength range'' that is optimal for constraining the stellar flux. The relative contribution from the planet to the total observed flux grows with wavelength, extending a lever-arm with which to isolate and constrain the planet spectrum and atmospheric properties. 

The lower panels of \Cref{fig:pie_spectrum} demonstrate how the PIE signal can be uniquely isolated when the stellar spectrum is properly accounted for (lower left panel), which allows the planet's emission spectrum to be fit using an atmospheric retrieval model and enables the inference of planetary atmospheric properties, such as the temperature structure (lower right panel). In this example, the broad wavelength coverage and high precision simulated JWST observations permit a robust inference of the TP profile. 

\begin{figure*}[!t]
\centering
\includegraphics[width=0.95\textwidth]{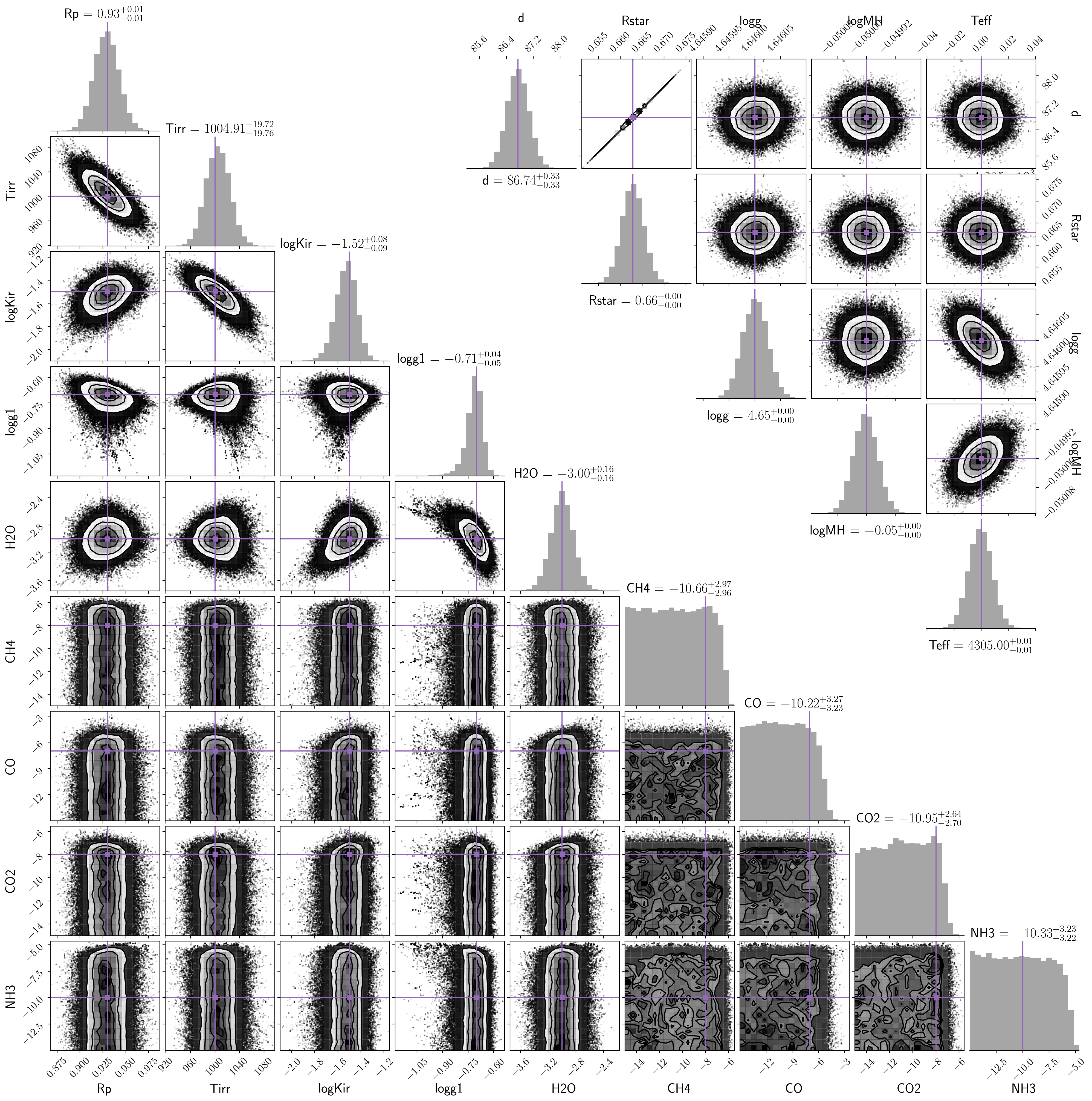}
\caption{Corner plot showing the 1D and 2D marginalized posterior distributions for the stellar (upper right) and planetary (lower left) parameters from a PIE retrieval. These results used data and uncertainties simulated from 4-hour exposures with each of JWST's NIRISS SOSS, NIRSpec G395, and MIRI LRS instrument modes. We identify no significant correlations between the stellar and planetary parameters (covariances are not shown).  Outside of the stellar radius/distance degeneracy, the stellar parameters are precisely constrained by the data.  For WASP-43b, the data are sufficiently precise to constrain the planet radius, nightside thermal structure, and \ce{H2O} abundance.  Additional telescope time is needed to adequately constrain the abundances of other prominent molecules.
}
\label{fig:double_corner1}
\end{figure*}

\Cref{fig:double_corner1} expands on the results of the retrieval shown in \Cref{fig:pie_spectrum} by showing the full 1D and 2D marginalized posterior distributions for the stellar and planetary parameters in the upper right corner and lower left corner, respectively. In general, posterior constraints on the stellar parameters are very tight. In practice, we anticipate that stellar spectral and temporal variability due to granulation, spots, and faculae will limit the precision of these inferred stellar parameters, but we discuss this further in Section \ref{sec:discussion}. As expected, the stellar radius and distance to the system are mathematically degenerate and show Gaussian posteriors that are identical to their respective priors taken from Gaia. Although \Cref{fig:double_corner1} omits covariances between stellar and planetary parameters, no such degeneracies were identified.  

The planetary parameters are also quite well constrained. The irradiation temperature is inferred to about $\pm 20$ K. The \ce{H2O} abundance is retrieved to a precision of $\pm 0.16$ dex. Although the abundances of \ce{CH4}, \ce{CO}, \ce{CO2}, and \ce{NH3} are not well constrained, we retrieve credible upper limits that are consistent with their respective low abundance input values. The planet radius and irradiation temperature show a negative correlation analogous to the findings of \citetalias{Stevenson2020} using simple blackbody models. Slight covariances are apparent between the parameters controlling the TP profile, but the TP profiles shown in \Cref{fig:pie_spectrum} calculated from posterior samples reveal an accurate picture of the input atmospheric thermal structure. 

We used the Bayesian Information Criterion (BIC) to verify that the JWST quality data warrants the inclusion of the planetary flux component in our retrievals. To compare against the results shown in \Cref{fig:pie_spectrum} and \Cref{fig:double_corner1}, we ran another retrieval on the simulated spectrum that omitted the planetary flux model and all of the corresponding parameters. We found that this yielded a $\Delta \mathrm{BIC} \approx 40,000$, favoring the model with components for the planet and star. This indicates that the fit to the spectrum is drastically improved by the addition of the planet model, which makes negligible the ``Occam's razor'' penalty incurred to the BIC by nine additional free parameters. Thus, it may be possible to use this type of model comparison to accurately diagnose the presence of PIE using precise, broad wavelength stellar spectra.  

In the following subsections we expand our investigation to consider wavelength range, non-transiting planets, exposure time, and exozodiacal light contamination.    

\subsection{Wavelength Range \& JWST Instrument Selection} 
\label{sec:results:instruments}

Our initial retrieval used simulated JWST data for 4-hour out-of-transit observations with NIRISS SOSS, NIRSpec G395, and MIRI LRS to achieve continuous wavelength coverage between $0.6 - 12$ {\micron}. Now we examine the result of using individual instruments. Table \ref{tab:results_free} shows our retrieval results where each column lists $1{\sigma}$ constraints on each retrieved model parameter (rows) for multiple experiments using different JWST instrument modes (columns). Note that the final column on the right shows results for 4-hours with all three instruments and corresponds to the results shown in \Cref{fig:pie_spectrum} and \Cref{fig:double_corner1}, while the other columns correspond to 1-hour exposures. 

As expected, retrievals on the full wavelength range provide the best constraints on the stellar and planetary parameters, but the performance of individual instruments varies from parameter to parameter. For example, we find that shorter wavelength observations that resolve the peak of the stellar SED offer the best constraints on the stellar parameters, with the tightest constraints on the effective stellar temperature ($T_{eff}$) provided by NIRISS, then NIRSpec, then MIRI. Similarly, the planet's 1000~K nightside emission peaks at 2.9 {\micron}; therefore, its irradiation temperature ($T_{irr}$) is best constrained by either NIRISS or NIRSpec, while MIRI offers a $3 \times$ worse constraint. However, the planet radius ($R_p$) is essentially unconstrained by the individual use of NIRISS or NIRSpec (where the posteriors are consistent with the priors), but shows an improved posterior constraint using only MIRI data. Despite the lack of radius constraint from using NIRISS or NIRSpec on their own, the planet radius is significantly better constrained using data from all three instruments. This result is consistent with the presence of a correlation between planet radius and irradiation temperature shown in \Cref{fig:double_corner1}, which we will revisit in Section \ref{sec:results:transit_vs_nontransit} for non-transiting planets.  Regarding constraints on \ce{H2O}, NIRSpec G395 provides the best individual constraint of $\pm 1.7$ dex, MIRI LRS provides the next best constraint of $\pm 2.3$ dex, and NIRISS SOSS is much worse at $\pm 5.1$ dex. These abundance results that favor the use of NIRSpec reflect a combination of the rising sensitivity to the planet thermal emission with wavelength and the specific wavelengths that exhibit \ce{H2O} absorption features. 

\begin{deluxetable*}{lrl|rrrrr}
\tablewidth{0.98\textwidth}
\tabletypesize{\normalsize}
\tablecaption{Retrieved constraints on WASP-43b using multiple JWST instruments to simulate the PIE technique \label{tab:results_free}}
\tablehead{\colhead{} & \colhead{} & \colhead{} & \colhead{NIRISS SOSS} & \colhead{NIRSpec G395} &  \colhead{MIRI LRS} &  \colhead{All} & \colhead{All} \\
           \colhead{Parameters} & \colhead{Defaults} & \colhead{Priors} & \colhead{(1 hr)} & \colhead{(1 hr)} &  \colhead{(1 hr)} &  \colhead{(1 hr)} &  \colhead{(4 hr)} }
\startdata
      $T_{eff}$         &   $4305.0$ & $\mathcal{N}(4305, 174.75)$   &    $\pm 0.020$ &    $\pm 0.065$ &     $\pm 0.16$ &     $\pm 0.018$ &    $\pm 0.0086$ \\
     $\log([M/H])$      &    $-0.05$ & $\mathcal{N}(-0.05, 0.17)$    & $\pm 0.000090$ &  $\pm 0.00080$ &   $\pm 0.0018$ &  $\pm 0.000069$ &  $\pm 0.000034$ \\
      $\log(g)$         &    $4.646$ & $\mathcal{N}(4.646, 0.052)$   & $\pm 0.000054$ &  $\pm 0.00031$ &  $\pm 0.00069$ &  $\pm 0.000050$ &  $\pm 0.000025$ \\
     $R_s$              &   $0.6629$ & $\mathcal{N}(0.6629, 0.0554)$ &   $\pm 0.0025$ &   $\pm 0.0024$ &   $\pm 0.0025$ &    $\pm 0.0025$ &    $\pm 0.0025$ \\
         $d$            &  $86.7467$ & $\mathcal{N}(86.75, 0.33)$    &     $\pm 0.33$ &     $\pm 0.32$ &     $\pm 0.32$ &      $\pm 0.32$ &      $\pm 0.33$ \\
        $R_p$           &     $0.93$ & $\mathcal{N}(0.93, 0.08)$     &    $\pm 0.080$ &    $\pm 0.079$ &    $\pm 0.068$ &     $\pm 0.026$ &     $\pm 0.014$ \\
      $T_{irr}$         &     $1000$ & $\mathcal{U}(300, 3000)$      &      $\pm 100$ &      $\pm 105$ &      $\pm 303$ &        $\pm 41$ &        $\pm 20$ \\
    $\log(\kappa_{IR})$ &     $-1.5$ & $\mathcal{U}(-3, 0)$          &     $\pm 0.89$ &     $\pm 0.44$ &     $\pm 0.78$ &      $\pm 0.18$ &     $\pm 0.090$ \\
     $\log(g_1)$        &     $-0.7$ & $\mathcal{U}(-3, -1)$         &      $\pm 1.0$ &     $\pm 0.83$ &      $\pm 1.1$ &      $\pm 0.18$ &     $\pm 0.044$ \\
       \ce{H2O}         &     $-3.0$ & $\mathcal{U}(-15, 0)$         &      $\pm 5.1$ &      $\pm 1.7$ &      $\pm 2.3$ &      $\pm 0.39$ &      $\pm 0.16$ \\
       \ce{CH4}         &     $-8.0$ & $\mathcal{U}(-15, 0)$         &      $\pm 4.9$ &      $\pm 3.6$ &      $\pm 5.1$ &       $\pm 3.1$ &       $\pm 3.0$ \\
        \ce{CO}         &     $-7.0$ & $\mathcal{U}(-15, 0)$         &      $\pm 5.0$ &      $\pm 4.7$ &      $\pm 5.1$ &       $\pm 3.7$ &       $\pm 3.3$ \\
       \ce{CO2}         &     $-8.0$ & $\mathcal{U}(-15, 0)$         &      $\pm 4.6$ &      $\pm 3.5$ &      $\pm 5.0$ &       $\pm 2.9$ &       $\pm 2.7$ \\
       \ce{NH3}         &    $-10.0$ & $\mathcal{U}(-15, 0)$         &      $\pm 5.1$ &      $\pm 3.8$ &      $\pm 5.3$ &       $\pm 3.5$ &       $\pm 3.2$ \\
\enddata
\tablecomments{$\mathcal{U}$ denotes a uniform prior probability distribution described in terms of the lower and upper bounds, and $\mathcal{N}$ denotes a normal prior described by the mean and standard deviation.}
\end{deluxetable*}

To examine how marginalizing over uncertainties in the stellar spectrum decreases the precision on retrieved planetary parameters when using the PIE technique, we ran a series of retrievals with the stellar flux fixed at the true input value. This simulates the traditional secondary eclipse retrieval approach. We kept the exposure time per instrument fixed at 4 hours and simulated individual cases for NIRISS, NIRSpec, and MIRI. Using a fixed stellar flux resulted in tighter constraints on, in particular, the planet's irradiation temperature and radius, with the largest effect seen towards longer wavelengths. When marginalizing over the stellar fit using the PIE technique we observed a $1.0\times$, $2.4\times$, and $3.5\times$ decrease in the precision on the retrieved irradiation temperature using NIRISS, NIRSpec, and MIRI, respectively, and a $1.1\times$, $2.3\times$, and $2.5\times$ decrease in the precision on the retrieved planet radius. We observed no decrease in the precision of molecular abundances. Thus, long wavelength PIE observations in the MIR that lack the NIR reference wavelength range to constrain the stellar spectrum suffer the greatest loss in precision of retrieved planetary information relative to traditional secondary eclipse measurements. 

One potential caveat that must be considered when applying the PIE technique to absolute fluxes is that JWST's absolute flux calibration accuracy may limit the ability to use the PIE technique on panchromatic spectra stitched together from observations using different JWST instruments with non-negligible offsets caused by their absolute flux calibration. We investigated this in detail in \autoref{sec:appendix:calibration} and found that if JWST achieves its nominal requirement of $2\%$ absolute flux calibration accuracy, offset correction parameters can be used within the PIE retrieval to enable the analysis of spectra obtained with different instruments. Using $2\%$ offsets between instruments, we find that the retrieved molecular abundances in the planetary atmosphere are completely unaffected, and a relatively small loss in precision (the one sigma uncertainties grow by $< 50 \%$) is incurred for the stellar radius, planetary radius, and irradiation temperature relative to the baseline results using all three instruments without calibration offsets (shown in \Cref{tab:results_free}). Even with $2\%$ instrument offsets to simulate absolute flux calibration, our results indicate that using all three JWST instruments---NIRISS SOSS, NIRSpec G395, and MIRI LRS–––to observe near-continuous wavelength coverage between $0.6 - 12$ {\micron} still outperforms any single instrument for nightside characterization of WASP-43b with the PIE technique. 

\subsection{Transiting versus Non-Transiting Planets} 
\label{sec:results:transit_vs_nontransit}

By using spectral information to separate planet from star, the PIE technique offers an opportunity to study the thermal emission from non-transiting planets. For the purposes of this work, we assume that non-transiting planets have known orbits but an unknown radius, like those detected from radial velocity (RV) surveys. However, we remain focused on a WASP-43b-like exoplanet.
Instead of using the Gaussian prior on planet radius shown in \Cref{tab:results_free} from transit observations, we use an uninformative uniform prior given by $\mathcal{U}(0.5, 1.5)$.  

\begin{figure}[t!]
\centering
\includegraphics[width=0.47\textwidth]{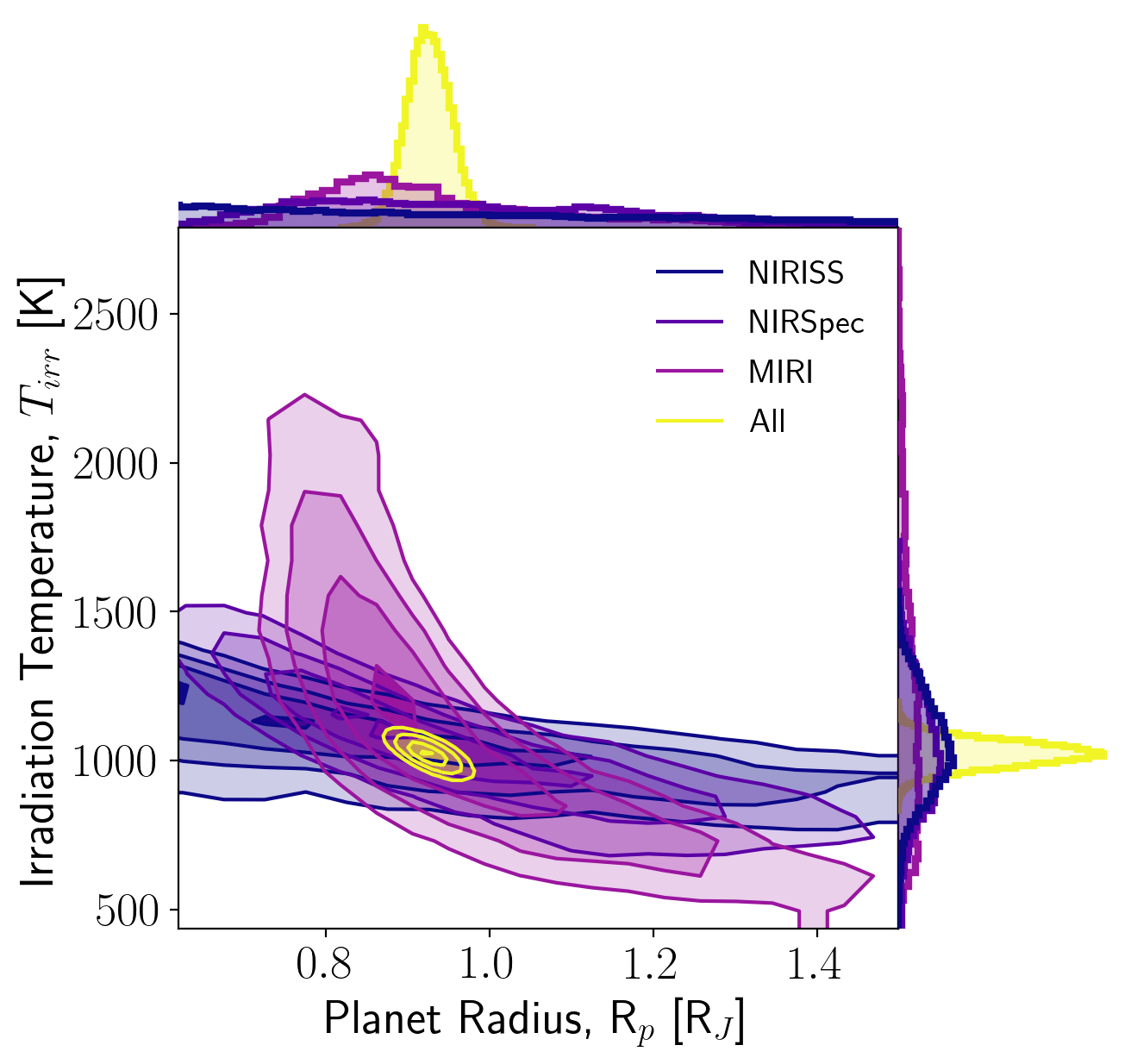}
\includegraphics[width=0.47\textwidth]{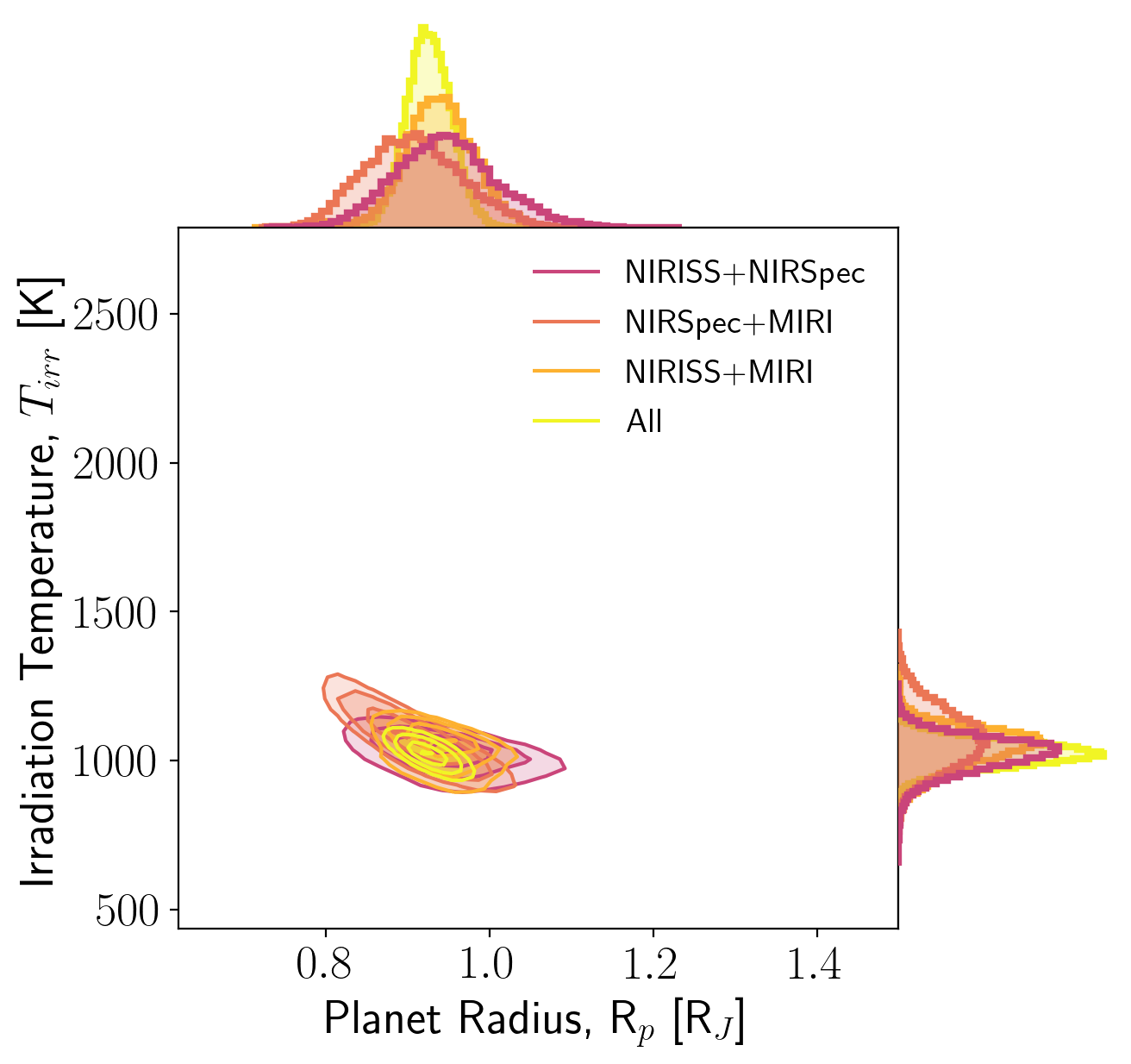}
\caption{One- and two-dimensional marginalized posterior distributions for the retrieved irradiation temperature and radius of WASP-43b from synthetic nightside observations with JWST. Observations assume 1 hour exposures with each instrument given in the legend. The degeneracy between planet temperature and radius depends on the observed wavelength range, and can be broken with sufficiently broad wavelength coverage due to the wavelength dependence (independence) of thermal emission on temperature (radius).} 
\label{fig:Tirr_vs_Rp}
\end{figure}

\Cref{fig:Tirr_vs_Rp} shows 1D and 2D marginalized posterior distributions for WASP-43b nightside PIE retrievals assuming no informative prior information about the planet radius. 
The 1D marginal posteriors show that, if only one instrument is used, planet temperature can be better constrained with shorter wavelength observations and planet radius can be better constrained at longer wavelengths. Both the planet radius and temperature can be much better constrained by combining data from two or three instruments. Using only two instruments, the irradiation temperature is best constrained with combined data from NIRISS and NIRSpec, while the planet radius is best constrained with NIRISS and MIRI. Intriguingly, the 2D posterior projection between irradiation temperature and planet radius shows how their covariance changes with each instrument. 

\subsection{Varying Exposure Time} 
\label{sec:results:exptime}

Here we briefly consider the effect of varying the exposure time. We repeated our previous retrieval that used data from 1-hour exposures with all three JWST instrument modes, now using 0.25-hour and 4-hour exposures to halve and double the signal-to-noise ratio (SNR) of the data, respectively, compared to the fiducial case. \Cref{tab:results_free} provides the constraints for the 1-hour and 4-hour cases.  
With the exception of prior-dominated parameters (e.g. the stellar radius and distance) and unconstrained parameters with non-Gaussian posteriors (e.g. \ce{CH4}, \ce{CO}, etc.), our exposure time experiment yielded results that scaled with SNR exactly as expected. For example, the $1 \sigma$ constraint on the planet radius dropped (rose) from $\pm 0.026$ R$_J$ to $\pm 0.014$ R$_J$ ($\pm 0.051$ R$_J$), approximately a factor of 2, for a $2 \times$ increase (decrease) in data SNR. We note that these tests do not include the effect of a noise floor in the observations, which may cause wavelength-dependent deviations in the scaling of SNR with the square-root of the exposure time that could propagate through the parameter inference. The effects of various noise floors on PIE retrievals will be explored in an upcoming paper.  
Our 4-hour NIRISS SOSS spectrum reaches a maximum precision of 17 ppm at 1.4 \micron{} with $\mathcal{R}=100$. 
While the exact wavelength coverage needed to precisely constrain the planet parameters will change with planet temperature, for the non-transiting exoplanet case we recommend maximizing the wavelength coverage instead of repeated observations with a specific instrument mode. 

\subsection{Sensitivity to Exozodiacal Dust}
\label{sec:results:zodi} 

A key concern for the PIE technique is the presence of exozodiacal dust in exoplanetary systems and how the thermal emission from such dust may be incorrectly attributed to planetary thermal emission and could propagate biases into the retrieval of atmospheric parameters. To estimate the magnitude of this effect within the context of our previous findings, we conducted a series of retrieval simulations with exozodiacal emission injected into the observed spectrum. We varied the intensity of the exozodiacal signal injected in the simulated data, but initially we did not attempt to retrieve the exozodiacal signal; instead we seek to characterize the bias that manifests in the inference if exozodiacal dust is not accounted for by the retrieval model. 

We modeled exozodiacal flux (emitted and reflected) in the WASP-43 system using \texttt{Zodipic} \citep{Kuchner2012}. We assumed the following properties for WASP-43, $R_*$=0.663\,$R_\sun$, $L_*$=0.136\,$L_\sun$, $T_*$=4305\,K, $d$=86.7467\,pc, $\log g$=4.492, and solar metallicity. By default, the dust in the disk is considered to scatter isotropically and scattered light is only considered when $\lambda<$ 4.2\,\micron{}. The inner spherical dust radius corresponds to a dust temperature of 1500\,K, which is appropriate for removal from the system via sublimation in the absence of a planet. In reality, a planet could change the morphology of the disk and cause dust to collect at various mean-motion resonances and clear out annuli \citep{Stark2013}, but we assume the simple case of an isolated and smooth disk out to 3.28\,AU. We modeled the spectral contribution from 0.3\,\micron{} -- 30\,\micron{} assuming a pixel size of 1\,mas.

For this exozodiacal sensitivity study, we considered only the case with the highest quality data, which used all three JWST instruments to span 0.6--12 {\micron} and assumed a 4-hour exposure time for each instrument mode. We tested adding 1 zodi, consistent with the Solar System level of zodiacal dust, 3 zodis, consistent with the median zodi level for Sun-like stars (\citealp{Ertel2020} found the median, $m=3_{-3}^{+6}$), and 10 zodis.

\begin{figure*}[t!]
\centering
\includegraphics[height=3.5in]{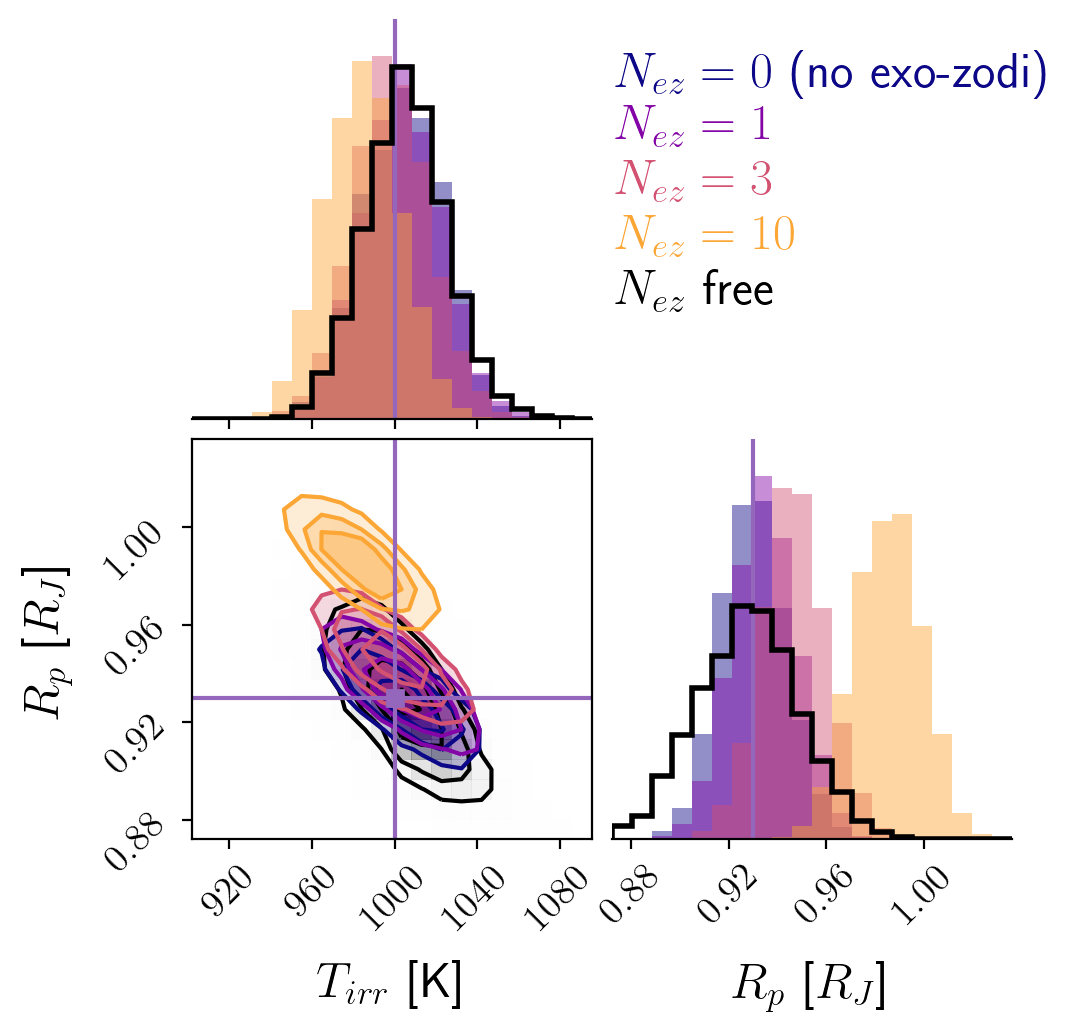}
\includegraphics[height=3.5in]{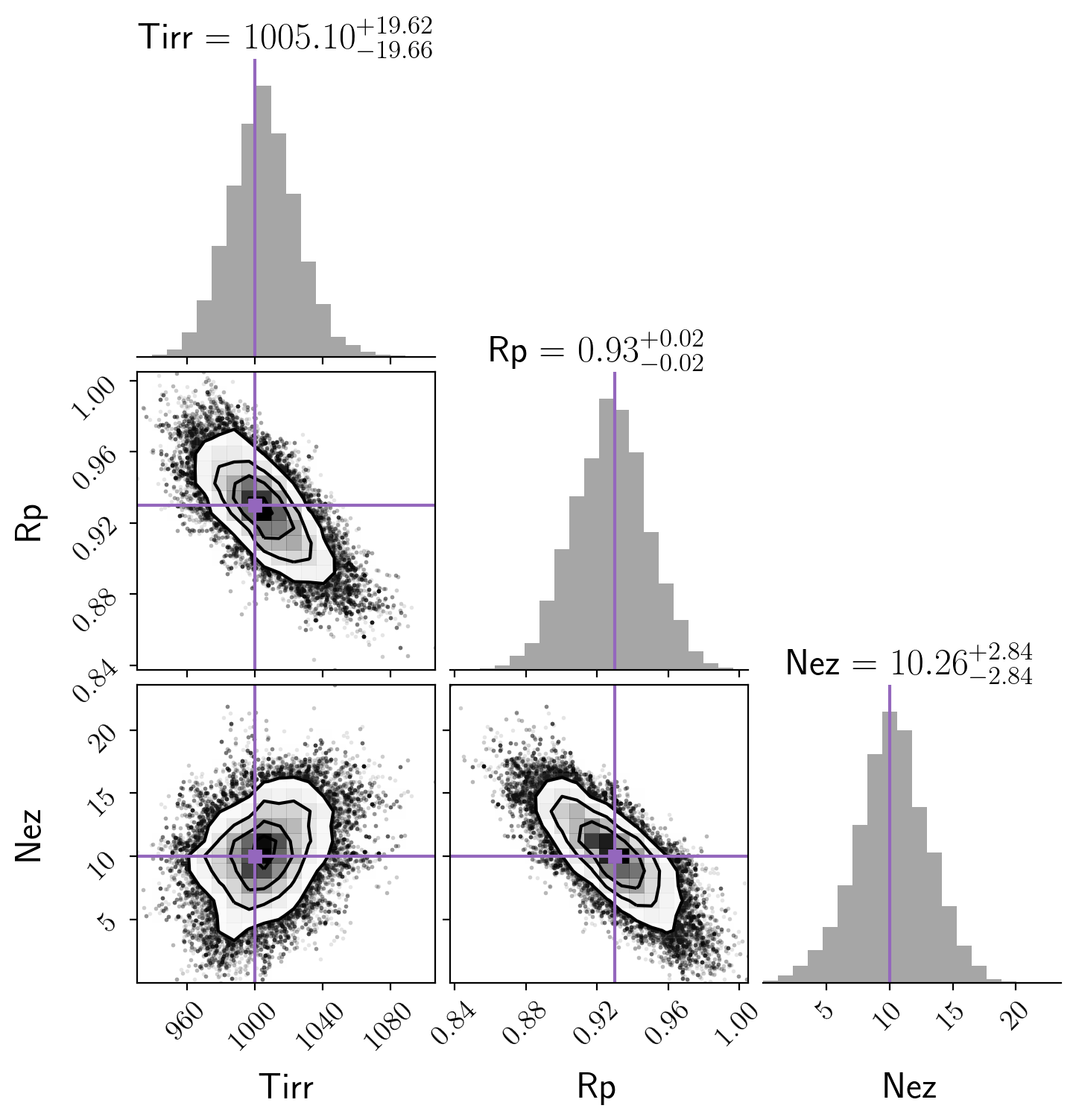}
\caption{\textit{Left:} One- and two-dimensional marginalized posterior distributions for the retrieved nightside irradiation temperature and radius of WASP-43b from synthetic JWST observations assuming four different levels of exozodiacal contamination. Failing to account for exozodiacal emission can bias the retrieved planet radius, and the planet's temperature to a lesser extent, particularly for systems with high exozodiacal levels. \textit{Right:} Marginalized posteriors for a PIE retrieval that explicitly fits for $N_{ez}$. The synthetic JWST data were generated with $N_{ez} = 10$. A subset of the total parameter space is shown to highlight the most affected parameters. Fitting for the zodi level removes the bias in planet parameters, but increases their uncertainties.} 
\label{fig:zodi}
\end{figure*}

We find that neglecting exozodiacal emission in PIE retrievals can bias the retrieved planet radius and, to a lesser extent, the planet temperature for systems with high exozodiacal levels. The left panel of \Cref{fig:zodi} shows the retrieved 1D and 2D marginalized posteriors for the planet radius and irradiation temperature for a set of simulations with varying exozodiacal contamination in the observed spectrum. For the case without exozodiacal contamination ($N_{ez} = 0$), we retrieve posteriors that are centered at the input value used to generate the JWST spectra. Increasing the exozodiacal contamination in the simulated system biases the retrieved planet radius to higher values in order to account for the additional system flux. This radius bias corresponds to 0.00558 $R_J$ (${\sim} 400$ km) per zodi in the WASP-43 system. Similarly, the retrieved irradiation temperature shifts to lower values with increasing zodi, consistent with the anticorrelation seen between radius and temperature. We find that the bias in radius can be statistically significant with $1.1 \sigma$ and $4.2 \sigma$ radius increases for $N_{ez} = 3$ and $N_{ez} = 10$, respectively, but the corresponding bias in irradiation temperature is negligible ($< 1 \sigma$ for $N_{ez} = 10$) for this JWST test case. We note that for our high exozodiacal case with $N_{ez} = 10$, the stellar gravity $\log(g)$, TP profile parameter $\log(g_1)$, and \ce{H2O} abundance are all marginally biased by about $1 \sigma$. 

As a final test of exozodiacal contamination, we incorporated a simple exozodiacal emission model into our PIE retrieval in order to explore the ability to mitigate the aforementioned biases. In this case, Equation \ref{eqn:system_flux} becomes 
\begin{equation}
\label{eqn:system_flux_zodi}
    F_{sys} = F_{s} + F_{p} + F_{ez}, 
\end{equation}
where $F_{ez}$---the exozodiacal spectrum---replaces $F_{\epsilon}$. The exozodiacal light is assumed for simplicity to be dependent only on the level of exozodiacal dust in the system, $N_{ez}$, which linearly scales the fiducial spectrum. We then performed the same retrieval experiment as above with $N_{ez} = 10$ contamination, except with $N_{ez}$ as an additional free parameter.

The right panel of \Cref{fig:zodi} shows the results of our exozodiacal retrieval for a subset of the free parameters that are most tied to the effects of exozodiacal emission. The zodi level in the system is constrained to $N_{ez} = 10.26 \pm 2.84$, which is consistent with our injected value. 
The covariance between planet radius and exozodiacal level shows a clear anticorrelation, but no significant correlation is seen between zodi and irradiation temperature. Including $N_{ez}$ as a free parameter successfully mitigates the radius bias; however, this comes at the cost of additional radius uncertainty due to the marginalization over the zodi uncertainty propagated through the radius-zodi covariance. Compared to the case with no exozodiacal light ($N_{ez} = 0$), the planet radius uncertainty in this scenario grows by a factor of ${\sim} 1.5$. Although we achieve a slightly better fit to the spectrum when we include $N_{ez}$ as a free parameter, using the BIC for model selection marginally favors the model without $N_{ez}$ ($\Delta \mathrm{BIC}=1.7$) due to the penalty of adding another dimension to the inference. This suggests that it may be difficult, in practice, to confidently diagnose the presence of exozodiacal contamination and mitigate potential biases, but systems with even greater zodi or a different system configuration may not have this issue. 

\section{Discussion \& Conclusion} \label{sec:discussion}

We have developed a novel exoplanet atmospheric retrieval algorithm for use with the PIE technique that performs a joint fit to unresolved stellar and planetary spectra in the near-to-mid IR. We used synthetic JWST observations of the hot Jupiter WASP-43b just outside of transit to explore the ability to use PIE to infer atmospheric characteristics of the planet's nightside. Using a simple five-parameter stellar spectral model and a nine-parameter planetary model, we find that the stellar and planetary spectra can be uniquely constrained with no apparent degeneracies between their respective parameters. We also find that typical out-of-transit observation times (e.g., 1--4 hours) may be sufficient to constrain planet properties. Broad wavelength coverage with JWST using a combination of NIRISS, NIRSpec, and MIRI observations will greatly benefit the inference of planetary atmospheric parameters using the PIE technique as long as JWST meets its nominal $2\%$ absolute flux calibration accuracy requirement. 

For non-transiting planets without well-constrained radii, the planet radius and (irradiation) temperature are degenerate. However, we find that the morphology of the degeneracy depends on the wavelength range of the data relative to the peak of the planetary emission spectrum. As a result, the PIE technique appears capable of breaking the radius-temperature degeneracy and constraining both parameters if sufficiently-broad wavelength coverage is available.  While we have not considered the mass-inclination degeneracy that exists for RV planets \citep{Hatzes2016}, the thermal signature of an exoplanet observed with the PIE technique, and its variability with time, will depend on orbital inclination and may enable it to be constrained. Furthermore, the radius constraints from PIE reported in this work could be combined with exoplanet mass-radius relationships to help break the mass-inclination degeneracy for RV planets. Future work will investigate the effects of unknown inclination on PIE. 

We investigated the potential for exozodiacal emission from dust grains in exoplanet systems to compromise the PIE technique and subsequent planet inferences. We found that, indeed, strong ($N_{ez} \sim 10$) unresolved exozodiacal emission in observed systems can cause a significant bias in the retrieved planet radius, and to a lesser extent planet temperature, which could be significant for non-transiting planets in systems with above average exozodiacal levels if left unaccounted for. However, by implementing a simple exozodiacal model within our PIE retrieval framework, we demonstrated that the bias can be mitigated, at a small cost of planet radius precision, by fitting for the zodi emission.

\subsection{Future Work}

Although we have demonstrated that the PIE technique may be a viable strategy for thermal studies of transiting and non-transiting hot exoplanet atmospheres, more work will be needed to expand on our findings and to relax assumptions that we have made (e.g. static stellar and planetary flux sources, single planet chemical composition, no systematic noise, etc.). Most critically, we have assumed a simple stellar spectral model that does not account for stellar granulation, spots, or faculae, which will all add complexity to the stellar spectrum inference problem that must be solved for any planet information to be inferred. A more sophisticated stellar model with multiple spectral components for cool spots and hot faculae may cause degeneracies in the stellar fit that could decrease the precision on inferred stellar parameters from those found in this work. It remains unclear if these stellar complexities will be degenerate with the planetary spectrum and parameters since their temperatures differ sufficiently well to be separated. 
Additionally, we have considered only a WASP-43b-like planet with a ${\sim}1000$ K nightside. The PIE technique will only increase in difficulty towards smaller and cooler planets that emit less flux. More work is needed across a broad range in planet and star temperatures to identify the optimal and plausible bounds for the PIE technique with JWST and other future mission concepts. 

Finally, our exploration into exozodiacal contamination highlights how additional unresolved flux sources may confuse and bias the PIE technique. We showed that increasing the complexity of our PIE retrieval model to account for exozodiacal light did mitigate the bias, but in the same stroke decreased the precision of our inferred planet radius. This serves as a cautionary warning if too many unresolved sources and/or yet unidentified complicating effects enter the picture: the fidelity of planetary information could be lost to these degeneracies. These questions and others require additional study in order to advance the maturity of the PIE technique. Ultimately, our work has demonstrated that the out-of-transit baseline observations that are already planned with JWST for many hot Jupiters will provide an opportune test bed for nightside atmospheric characterization using the PIE technique.  

\acknowledgments

We thank the anonymous referee for their thoughtful comments that helped improve the quality of this manuscript. We also thank M. Line for developing and maintaining the open-source CHIMERA code that made this work possible. 
This material is based upon work supported by the National Aeronautics and Space Administration (NASA) under Grant No. 80NSSC21K0905 issued through the Interdisciplinary Consortia for Astrobiology Research (ICAR) program.
This work was also funded by internal research and development funding from the Johns Hopkins Applied Physics Laboratory. 

\software{Astropy \citep{Astropy2013, Astropy2018}, CHIMERA \citep{Line2013a, Line2014}, corner \citep{corner}, emcee \citep{Foreman-Mackey2013}, Matplotlib \citep{Hunter2007}, NumPy \citep{Walt2011}, SciPy \citep{Virtanen2019scipy}, Pandas \citep{pandas2010}, Pandeia \citep{Pontoppidan2016}, PandExo \citep{Batalha2017b, Pandexo2018}, pysynphot \citep{STScI2013}, Zodipic \citep{Kuchner2012}}

\bibliography{ms}

\appendix

\section{Absolute Flux Calibration Accuracy} \label{sec:appendix:calibration} 

To investigate the possible effects of absolute flux calibration on our PIE retrievals of spectra collected with different JWST instruments, we conducted an additional experiment with offsets between the NIRISS, NIRSpec, and MIRI data. Given that the nominal JWST requirement is 2\% absolute flux prediction accuracy for these instruments\footnote{\url{https://jwst-docs.stsci.edu/data-processing-and-calibration-files/absolute-flux-calibration}}, we generated synthetic observations using our 4-hour exposure experiment from \Cref{fig:pie_spectrum} and \Cref{fig:double_corner1} with flux offsets imposed between the spectra of NIRISS, NIRSpec, and MIRI to misalign the panchromatic spectrum. We considered two different cases with $2\%$ calibration offsets: ``Case 1'' with $+2\%$, $-2\%$, and $+2\%$ flux offsets for NIRISS, NIRSpec, and MIRI, respectively, and then we flipped the signs of the offsets for ``Case 2'' and used $-2\%$, $+2\%$, and $-2\%$. To account for the absolute flux calibration in our retrieval, we added three flux offset parameters---one for each instrument---to allow the retrieval to fit for the proper offsets between the model spectrum and the data. We did not consider how slight overlaps in instrument wavelength range (e.g. between NIRSpec G395 and MIRI LRS) could be used as another means to correct for offsets due to calibration. We used uniform priors on the three flux offset parameters of $\mathcal{U}(-0.05, +0.05)$ to keep them bounded within ${\pm} 5\%$. We then ran three retrieval experiments, two using the Case 1 and Case 2 $2\%$ calibration offset spectra and a third ``Case 3'' using the original data with $0\%$ offsets to explore the impact of marginalizing over the three additional calibration parameters. 

\begin{figure*}[!b]
\centering
\includegraphics[width=0.95\textwidth]{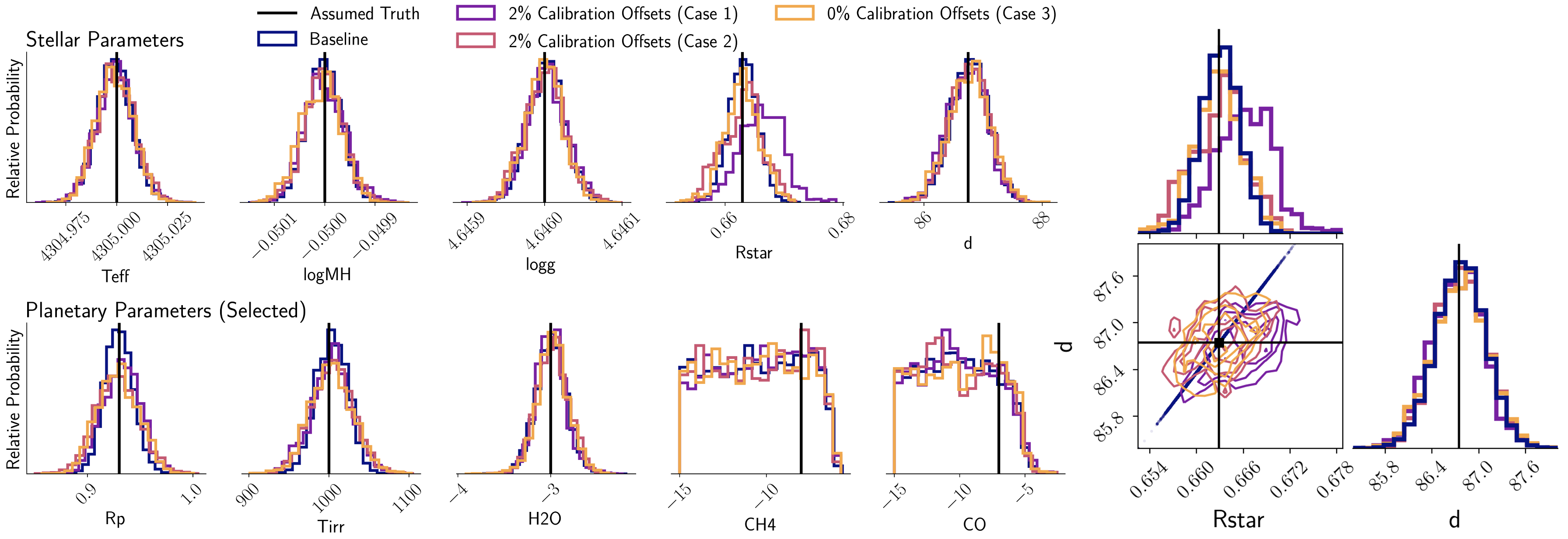}
\caption{Posterior PDFs for a subset of the retrieved parameters for our investigation into absolute flux calibration offsets between JWST instruments compared against the assumed input values (black lines) and baseline retrieval analysis. The top row of panels show the stellar parameters, the bottom row shows a selection of the planetary parameters, and the 2D contour on the right shows how the positive correlation between distance and stellar radius is broadened when instrument offset parameters are included in the PIE retrievals.} 
\label{fig:abs_flux_offset}
\end{figure*}

Our results show that, in general, accounting for the $2\%$ absolute flux prediction accuracy does reduce the retrieved precision on some of the stellar and planetary parameters, but the PIE method still yields reliable results that are a significant improvement over the results from any one instrument. \Cref{fig:abs_flux_offset} shows a subset of the retrieval results for our three calibration offset experiments compared against the baseline case with no instrument offsets and no offset correction parameters. The stellar radius is the most affected parameter by the $2\%$ offset cases, which yielded $50\%$ and $40\%$ larger retrieved uncertainties for Case 1 and Case 2, respectively, and is biased slightly high in Case 1. The stellar radius and distance degeneracy is shown in the right panel and appears significantly broadened for cases with offset corrections due to an anticorrelation between stellar radius and the offset parameters. The distance to the system is much less correlated with the offset parameters. The planet radius and irradiation temperature have $30 - 40\%$ larger uncertainties in these experiments, but the constraints on gas abundances are unaffected by our procedure to account for absolute flux calibration. These results indicate that if JWST achieves its nominal requirement of $2\%$ absolute flux calibration accuracy, the PIE technique will be capable of analyzing spectra obtained with different instruments with offsets between them, incurring a relatively small ($< 50 \%$) loss in precision for some retrieved planetary parameters. 

\end{document}